\documentclass[prl,twocolumn,aps,showpacs,preprintnumbers,amsmath,amssymb]{revtex4}
\usepackage{graphicx}
\usepackage{dcolumn}
\usepackage{bm}
\usepackage{amsmath} 
\begin{document}

\title{Splitting and merging an elongated Bose-Einstein condensate 
at finite temperature}
\author{A. Mebrahtu,$^{1,2}$, A. Sanpera,$^{1,3}$, and M. Lewenstein$^{1,4}$}
\affiliation{$^1$ Institut f\"ur Theoretische Physik, Universit\"at
  Hannover, D-30167 Hannover,Germany\\
$^2$ Department of Physics, Mekelle University, P. O. Box 231,
  Mekelle, Ethiopia.\\
$^3$ ICREA and Grup de F\'isica Te\`orica, Universitat Aut\`onoma de 
Barcelona, E-08193 Bellaterra, Spain.\\
$^4$ ICREA and ICFO-Institut de Ci\`encies Fot\`oniques, 
Parc Mediterrani de la Tecnolog\'ia, E-08860 Barcelona, Spain.}


\begin{abstract}
We analyze coherence effects during the splitting of a quasi one-dimensional 
condensate into two spatially separated ones and their subsequent merging 
into a single condensate. Our analysis takes into account finite-temperature 
effects, where phase fluctuations play an important role. We show that, at 
zero-temperature, the two split condensates can be merged into a single one 
with a negligible phase difference. By increasing temperature to a finite 
value below the critical point for condensation ($T_c$), i.e., 
$0 \le T/T_c < 1$, a considerable enhancement of phase and density 
fluctuations  appears during the process of splitting and merging. 
Our results show that if the process of splitting and merging is
sufficiently adiabatic, the whole process is quite insensitive to phase
fluctuations and even at high temperatures, a single condensate 
can be produced.
\end{abstract}


\maketitle

\section{Introduction} 
\label{intro}
The experimental exploration of Bose-Einstein
condensates (BECs) in many different systems such as alkali metal
gases~\cite{Anderson,Davis,Bradley}, hydrogen~\cite{Fried}, 
meta-stable helium~\cite{Robert}, and 
molecules~\cite{Jochim,Zwierlein03,Greiner03} 
has led to a surge of
interest in manipulating ultracold atomic samples under very 
different circumstances. One of the initial motivations for such
development was and is the prospect of creating a continuous and
coherent atomic beam~\cite{Koehl,Pinkse,Chikkatur}, which is the
atomic analogy of the optical laser beam~\cite{Maiman}.

Among the major challenges in achieving a continuous atom laser is
how to overcome the difficulty due to the restrictive cooling conditions 
for continuously condensing the atomic gas. Spatial separation
of the evaporative cooling from the destructive laser cooling may
play a role in alleviating these challenges~\cite{Yi}. Recently,
a continuous BEC source was achieved by periodically replenishing a 
reservoir with condensates~\cite{Chikkatur}. 
There, optical tweezers were used to
transport sodium condensates from where they were created to the
reservoir. Such procedure allows one to continuously replenish the
reservoir which contains on average more than $10^6$ atoms.
Bringing a new condensate to the reservoir each time implies the
merging of condensates with different relative phases since each
condensates is created with a different phase. The merging of two
condensates with different phases poses a limitation on the
coherence of the process. 

Recently, interference effects in the merging of 
30 uncorrelated Bose-Einstein 
condensates released from a one-dimensional optical lattice have been 
discussed in~\cite{Hadzibabic}, whereas coherent splitting of 
BECs by deforming an optical single-well into a double-well potential 
for a trapped atom interferometer have been addressed in 
Refs.~\cite{Shin,Pezze}. Very recently, Schumm {\it et al.}~\cite{Schumm} 
has demonstrated  a coherent beam splitter on an atom chip by 
splitting the condensate in a double-well potential and merging it again. 
They have demonstrated phase preservation in this
process, even in the case when the split BECs are far enough apart to
inhibit tunnel coupling.

In this paper, 
we analyze the axial splitting of a very much elongated 
cigar-shape condensate into two condensates~\cite{Javanainen,Leggett} 
and their subsequent merging along the axial direction. 
Our analysis includes finite-temperature effects. 
In other words, phase fluctuations arising due to temperature 
are taken into account during the whole process: splitting and merging. 
We observe that as long as the process of splitting and merging is 
adiabatic enough, both the split and the merged condensates, 
even at relatively large temperatures, 
do survive the effects of these fluctuations. 

Low-dimensional quantum gases exhibit very fascinating properties
and have attracted a lot of interest, both theoretically and
experimentally~\cite{Druten96,Petrov1D,Moritz,Stoeferle,Paredes,Kinoshita,Goerlitz}. It is known that low-dimensional [one- (1D) and two- (2D) dimensional]
quantum gases differ qualitatively from their 3D 
counterparts~\cite{Petrov1D,Petrov2D,Kagan2D,Safonov}. From a theoretical
point of view, the use of a mean-field theory to describe a low-dimensional 
quantum gas is severely restricted. 
A widely used criterion to apply a mean-field approach is 
to demand that the average distance between particles, {\rm d}, is clearly smaller 
than the correlation length of the condensate 
$l_c=\hbar/\sqrt{m n g}$ where $m$, $g$, and $n$  
denote the mass, the interaction coupling, and the density, respectively.
In three dimensions, the above condition leads to $l_c/d\propto n^{-1/6}$ 
and is well satisfied for small densities, and the description of the 
system at $T=0$ with a mean-field Gross-Pitaevskii equation is fully 
satisfactory.
In the one-dimensional case, however, this ratio behaves as  
$l_c/d\propto n^{1/2}$ and this fact changes drastically the range of 
applicability of a mean-field treatment. 
A careful analysis of the validity of a mean-field
treatment in the 1D case~\cite{Pitaevskiibook} 
leads to the following condition:
\begin{equation}
\left( \frac{N a_s a_z}{a_\perp^2} \right)^{1/3}\gg 1,
\label{1DGP}
\end{equation}
where $N$ is the number of condensed atoms, 
$a_z=\sqrt{\hbar/(m \omega_z)}$ and $a_\perp=\sqrt{\hbar/(m\omega_\perp)}$ 
are the axial and radial oscillator lengths, respectively, and  $\omega_z$ 
and $\omega_\perp$ are the angular frequencies in the corresponding directions.
Thus, in 1D, contrary to the 3D case, when the density decreases 
the gas gradually becomes strongly correlated, acquires a fermionic 
character, and enters into the so-called Tonks-Girardeau 
regime~\cite{Girardeau,Menotti,Kheruntsyan,Pedri}. 
Experimental demonstration of a Tonks gas has been recently 
achieved~\cite{Paredes}.

The possibility of generating low-dimensional bosonic gases 
raises the question of the effects of quantum fluctuations. In an
untrapped 1D Bose system these fluctuations destroy finite- as
well as zero-temperature condensation. For trapped Bose gases, the 
situation is quite different: for noninteracting bosons in such a trap 
the finite spacing between the lowest and the next energy level allows for 
the occurrence of 1D Bose-Einstein condensation even at 
finite-temperatures as stipulated in Refs.~\cite{Druten96,Petrov1D}.
In such a case the absence of gapless excitations indicates that the  
BEC will not be destroyed immediately as interactions between bosons 
are turned on. 

In 1D trapping geometries, long-wavelength density and phase
fluctuations lead to a new intermediate state between a condensate
and a noncondensed system, which is commonly referred to as a
{\it quasicondensate}. In quasicondensates, the phase of the condensate
is only coherent over a finite distance that is smaller than the
system size. In other words, the phase coherence length is smaller
than the axial size of the sample. To understand the nature of 
quasicondensates at finite-temperature, one has to analyze the
behavior of the single particle correlation function by 
calculating the fluctuations of phase and density as has been 
done by Petrov {\it et al.}~\cite{Petrov1D}. There it is shown that for
temperatures below the degeneracy temperature, the condensate's
phase indeed fluctuates, but fluctuations of the density are 
still highly suppressed. This character of thermal fluctuations
is also present in highly elongated 3D gases~\cite{Petrov3D}, and has 
been recently observed experimentally 
in~\cite{Dettmer,Kreutzmann,Hellweg,Richard,Gerbier,Kadio}.

Quasi 1D geometries can be accessible in magnetic traps with a
cylindrically symmetric harmonic potential along the axial
direction that have transverse frequencies $\omega_{\perp}$  much
larger than the axial one $\omega_z$. In such configurations
the resulting condensate looks like a cigar aligned along the
symmetry or z axis.  With current technology, condensates with
aspect ratio $\lambda=\omega_{\perp}/\omega_z$ as large as 1000
are achievable.

In this paper, we study the process of splitting and subsequent 
merging of an elongated condensate confined in a 1D geometry
both at $T=0$, i.e., when the condensate has a well-defined phase,
and at finite-temperature ($T > 0$), in the quasicondensate regime.
For finite $T$, we analyze the process of splitting and merging 
for a wide range of temperatures, i.e., 
$T_{\phi} < T < T_d,$ where 
$T_{\phi}= T_d\hbar\omega_z/\mu$ ($\mu$ being the chemical potential) 
corresponds to the characteristic temperature above which a true-condensate 
turns into a quasicondensate in which phase fluctuations begin 
to play a role. On the other hand 
$T_d=N \hbar\omega_z/k_B$ is the 1D 
degeneracy temperature~\cite{Petrov1D}. 
The transition, or crossover, between the different regimes for the
1D degenerate interacting bosonic gas, i.e., true-condensate,
quasi-condensate, and Tonks gas is smooth. Thus, in the regime of
quasi-condensation density fluctuations are relatively suppressed while
phase fluctuations are enhanced. By keeping all parameters fixed and 
reducing the number of atoms, phase fluctuations 
become more and more pronounced, mean-field theory fails, and the gas 
enters into the strongly correlated regime
or Tonks gas~\cite{Girardeau,Menotti,Kheruntsyan,Pedri}.

In our case, the splitting of the condensate is achieved 
by means of a double-well potential grown adiabatically 
on top of a trapping harmonic potential. By adiabatically switching off the 
double-well potential, we merge these condensates into a single 
one (Merged BEC). We would like to stress that we use the mean-field 
Gross-Pitaevskii equation~(GPE) throughout this paper. It is worth pointing 
out that although the Gross-Pitaevskii equation~\cite{Dalfovo} 
describes properly coherent evolution of the atomic mean-field at $T=0$, 
it can also be used to solve time evolution at finite-temperature in a 
relatively straightforward manner. It might look like that the mean-field 
method of the GPE allows to make statements only about first order coherence. 
But, as it is well known, a closely inspection reveals that the GPE contains 
as such classical Bogoliubov-de Gennes~(BdG) equations, i.e. equations 
describing small fluctuations around a given solution of the GPE. 
Since we are here interested  in the study of relatively high 
temperatures $T > T_{\phi}$, classical description of fluctuations is fully 
appropriate. It turns out that one can simulate temperature effects
by adding fluctuations to the ground state solution of the GPE at $T=0$ 
in a way which mimics thermal fluctuations. At low temperatures 
$T < T_c$, where $T_c = N/\ln(2N)\hbar\omega_z/k_B$~\cite{Druten96}, 
this can be generally done by identifying phonon (quasi-particle) 
modes, i.e. eigensolutions of the BdG equations. The fluctuations are 
expressed thus as sums over the quasi-particles with amplitudes 
taken from Monte Carlo sampling and corresponding to the thermal (Boltzmann)
populations of the quasi-particle modes. Such a method is used in this paper, 
with the additional simplification that for quasi-1D  
situations only the phase fluctuations are relevant.

Note, that once we add the fluctuations initially at $t=0$, and as
long as they remain small in the course of evolution, they will
propagate in time with a very good approximation as appropriate
solutions of the time dependent BdG equations.  Note also, that our
approach allows in principle to obtain information about all
correlation functions: either in the form of an average over Monte Carlo
realizations of the initial fluctuations, or in the form of time
averages, due to the (expected) ergodic character of the evolution.
In some cases, even averaging over the initial data is not necessary:
the results for different realizations are so similar that looking for
few single cases allows one to draw conclusions about ``coherence''. 
By coherence we mean here, that the splitting process of an initial 1D 
BEC into two spatially separated 1D BECs occurs with a well-defined 
relative phase between them. Their subsequent merging into a single 
merged BEC, when the process is fully coherent, should result in a single 
condensate with a well-defined relative phase with respect to the initial one. 
If this is the case, there is a perfect overlap between the densities 
of the initial and merged condensates.  

It is worth stressing that our approach is a simplified version of the 
classical field methods used by several groups~\cite{Kadio,Davis01,Davis02,Davis01JPB,Davis03,Brewczyk,Goral01,Goral02,Schmidt}.
In their approach some emphasis was put on the explanation of  
the temperature concept, and a certain model of finite-temperature 
effects has been studied. Particularly interesting is here the 
possibility of extracting higher order correlations from a single shot 
measurements~\cite{BachPRA70,Altman,BachPRL92}, which in our case 
corresponds to a single realizations of the initial fluctuations.

\section{Description of the model}
\label{Sec2}
Here we consider $^{87}$Rb condensate with $N=1.2\times 10^4$
atoms confined in an harmonic trap with frequencies
$\omega_{\perp}=2\pi\times 715$ Hz and $\omega_z=2\pi\times 5$ Hz. 
For such parameters, the system is in the 1D 
Thomas-Fermi regime ($\mu \gg \hbar \omega_z$) along the axial direction.
The 1D chemical potential is given by 
\begin{equation}
 \mu = \hbar \omega_z\Bigg(\frac{3}{4\sqrt{2}} 
  \frac{N\, m\,g_{\rm 1D}\sqrt{\hbar/m\omega_z}}{\hbar^2}\Bigg)^{2/3}
\label{1Dmu}
\end{equation}
where $g_{\rm 1D}=2\hbar\omega_{\perp}a_s$ is the effective 1D coupling 
strength~\cite{Olshani}. For our parameters, transverse excitations are 
suppressed ($\mu \simeq \hbar \omega_{\perp}$), and the dynamics of such 
a Bose gas can be described by the usual mean-field GPE in 1D
\begin{eqnarray}
  i \hbar \frac{ \partial \Psi(z,t)}{\partial t} 
  & = & \Big(-\frac{\hbar^2}{2\,m} \frac{\partial^2}{\partial z^2}
       + V_{T}(z,t) \nonumber \\
  & + & g_{\rm 1D} N|\Psi(z,t)|^2 \Big) \Psi (z,t),
 \label{1D-GPE-T0}
\end{eqnarray}
where $\Psi (z,t)$ is the mean-field order parameter, or in other words the 
condensate wave function. The potential term $V_T(z,t)$ includes both the 
magnetic trap and the double-well potential as described below:
\begin{eqnarray}
  V_{T}(z,t) 
  & = & V_{\rm trap}(z) + V_{\rm op}(z,t) \nonumber \\
  & = & \frac{1}{2} m \omega_{z}^2 z^2 + 
  S(t) V_0 \,\cos(k_l \,z)^2.
  \label{Ext-pot-tot}
\end{eqnarray}

\begin{figure}[ht!]
  \begin{center}
    \includegraphics[width=\linewidth]{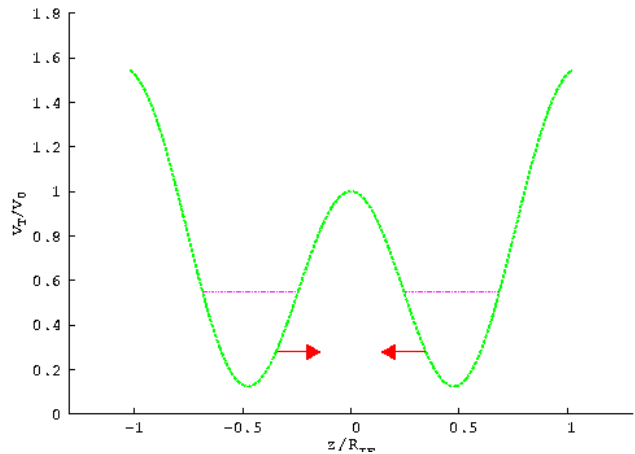}
    \caption{(Color online) Illustration of a double-well potential for
      splitting and merging of an elongated 1D BEC:
      a strong optical potential and an axial trapping potential 
      are combined for  creating a double-well potential which is 
      used for splitting a 1D condensate, when switched 
      on adiabatically, into two spatially separated symmetrical 
      1D condensates at the center of each well, and then merge 
      them into one when switched off adiabatically again.}
    \label{Pot}
  \end{center}
\end{figure}
The potential used to split the condensate into two spatially 
separated condensates, $V_{\rm op}(z,t),$ is switched on and off 
adiabatically by means of a time-dependent function $S(t).$  
The maximum depth of this potential is $V_0=2.2\times 10^4 E_r$ in 
terms of the recoil energy $E_r= \hbar^2 k_l^2/2m $ and $k_l=2\pi/\lambda_l.$ 
To achieve spatial separation of the split condensates one has to 
require that the distance between the two wells is, at least, of the order 
of the Thomas-Fermi radius ($R_{\rm TF}.$) This radius is 
given by $R_{\rm TF}=\sqrt{2\,\mu/(m\omega_{z}^2)},$ 
so that $k_l=\pi/R_{\rm TF}.$ For $N\simeq 10^4$ atoms, 
the Thomas-Fermi radius is $R_{\rm TF} \simeq 88 \mu$m. 
This is in agreement with the experimental results in Refs.~\cite{Shin,Schumm}.

The time dependent function $S(t)$ controls the switching on and off of 
the double-well potential and hence the overall splitting 
and merging process of the condensates. We define this function as
\begin{eqnarray}
  S(t) =\left\{\begin{array}{llll}
  & 0,  & {\rm for} \hspace*{2mm} t \le t_{\rm evo} \,{\rm and}\,  t \ge t_{\rm mer}, \\
  & \sin\Big(\frac{\pi}{2}\frac{t-t_{\rm evo}}{t_{\rm spl}-t_{\rm evo}}\Big)^2,
        & {\rm for} \hspace*{2mm} t_{\rm evo} < t < t_{\rm spl},           \\
  & 1,  & {\rm for} \hspace*{2mm} t_{\rm spl} \le t \le t_{\rm con}, \\
  & \cos\Big(\frac{\pi}{2}\frac{t-t_{\rm con}}{t_{\rm mer}-t_{\rm con}}\Big)^2,
        & {\rm for} \hspace*{2mm} t_{\rm con} < t < t_{\rm mer}.
  \end{array}
  \right.
  \label{tevo-tmer}
\end{eqnarray}
In this equation~(\ref{tevo-tmer}), $t_{\rm evo}$ is the time duration 
required for evolving the 1D BEC in real time before the 
splitting process begins. It extends for 10ms. Just immediately after, 
the splitting process begins and continues for a time interval $t_{\rm spl}$.
At the end  of $t_{\rm spl}$, the function $S(t)$ attains a maximum value of 
unity and remains constant for a time interval $t_{\rm con}$. During this time 
interval, two spatially separated 1D BECs are created. 
At the beginning of the time span $t_{\rm mer}$, merging of the condensates 
starts by switching off the double-well potential.
This process continues until the two 1D BECs merge together. 
A complete merging is only possible when  $S(t)$ becomes finally zero, i.e., 
when the optical potential is completely switched off. In this case the atoms 
remain only under the influence of the trapping potential. 

To ensure coherence during the process of the switching on and off of the 
optical potential, the raising of the double-well has to be slow enough to 
avoid excitations and to allow for quantum tunneling between the two wells.
Notice that the relevant time scale of excitations is given by the inverse 
of the frequency of the trap, in our case 
$t_{\rm sys}=2\pi/{\omega_z}=200$ ms. 
If $t_{\rm spl},t_{\rm mer}\gg t_{\rm sys}$ we expect coherent 
splitting and merging. On the other hand,  for $t_{\rm spl},t_{\rm mer}$ 
of the order of  $t_{\rm sys}$ or less, the process creates more and more 
excitations that cause incoherence. 
The faster the double-well potential switches on and off the stronger the 
excitations are. 
To check our claims we have carried out further numerical simulations 
for several values of $t_{\rm spl}$ and $t_{\rm mer}$ around $t_{\rm sys}$. 
In general, we have observed that a coherent process, at $T=0$,  
is achieved for switching times of 400 ms (or larger). 

\begin{figure}[ht!]
  \begin{center}
   \includegraphics[width=\linewidth]{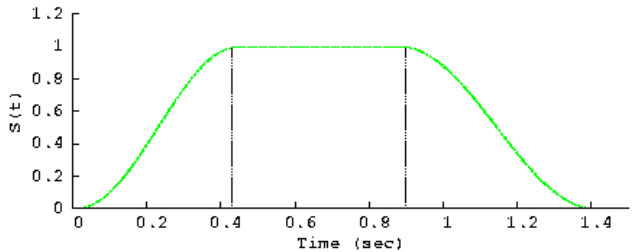}
  \end{center}
  \caption{(Color online) Time dependence of the function $S(t)$.
    For the whole process, we fixed a time evolution of 
    $t_{\rm evo}=10$ ms,
    a splitting time of $t_{\rm spl}=440$ ms, a time
    $t_{\rm con}=430$ ms in which $S(t)$ remains 
    constant (unity), and a time of 
    merging $t_{\rm mer}=530$ ms and an
    additional time of $10$ ms for allowing the merged condensate to make 
    its final evolution into a single condensate.}
\label{SplitParam}
\end{figure}

The full time dependency of $S(t)$ necessary for the coherent 
splitting and merging of the condensates based on Eq.~(\ref{tevo-tmer}) 
is illustrated in Fig.~\ref{SplitParam}.

\section{Splitting and merging at zero temperature}
\label{sec3}
We calculate first the ground state of the system by evolving the 
GPE~(\ref{1D-GPE-T0}) with $S(t)=0$ in imaginary time. 
After the ground state has been found, we numerically
solve Eq.~(\ref{1D-GPE-T0}) with the pulse profile given by the time 
dependent-function $S(t)$. 

\begin{figure}[ht!]
  \begin{center}
\includegraphics[width=\linewidth]{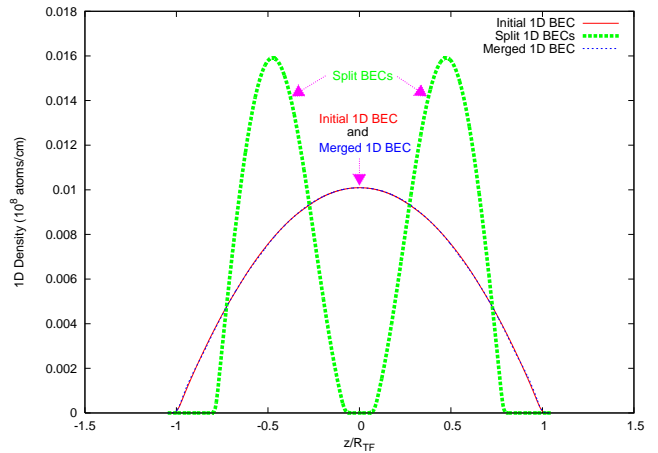}
 \end{center}
  \caption{(Color online) Adiabatic splitting and merging of an elongated 
    condensate at $T=0$. After coherently splitting an initial 1D 
    condensate (Initial BEC) into two spatially separated ones (Split BECs), 
    by adiabatically switching on the double-well potential given in 
    Fig.~\ref{Pot}, centered at $z/R_{\rm TF}=-0.5$ and $z/R_{\rm TF}=0.5$,  
    the optical potential is again switched off adiabatically. 
    This leads to a coherent merging of the Split condensates into a single 
    one (Merged  1D BEC). The merged  1D BEC overlaps on top of the 
    initial 1D BEC which is an indication of a completely coherent 
    merging process.}
  \label{Merg-densT0}
\end{figure}

The results of our simulations at temperature $T=0$, 
are summarized in Fig.~\ref{Merg-densT0}. There we display 
the condensates' density at three different times. First at $t=0,$ the  
initial condensate (Initial 1D BEC) has a characteristic Thomas-Fermi profile 
with $R_{\rm TF}=88 \mu$m.
Then at $t=800$ ms (corresponding to the ramping up of the double-well 
potential in 400 ms and keeping it constant during additional 400 ms),  
two spatially well separated identical condensates (Split BECs) appeared, 
centered at  $z/R_{\rm TF}=\pm 0.5$, each of them with a number of atoms 
$N_{s}=N/2$. 
Finally, we display the density of the condensate at $t=1400$ ms. The 
merged condensate (Merged 1D BEC) has exactly the same profile as the 
initial  one (Initial 1D BEC) and thus 
they cannot be distinguished in the figure, asserting that the process of 
splitting and merging is fully coherent. 

Coherence may be  a prerequisite for further applications such as in atom 
interferometry and quantum-information processing~\cite{Shin05}. 

\begin{figure}[ht!]
  \begin{center}
\includegraphics[width=\linewidth]{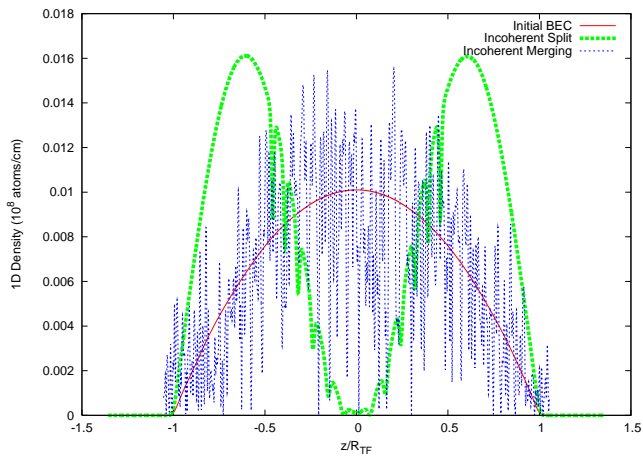}
 \end{center}
  \caption{(Color online) Nonadiabatic splitting and merging of an elongated 
    condensate at $T=0$. 
    The Initial BEC is split incoherently by considering a very small 
    splitting time~($t_{\rm spl}$) into two picked spatially 
    separated ones (Incoherent Split) by nonadiabatically 
    switching on the double-well potential~(Fig.~\ref{Pot}).  
    The optical potential is again switched off nonadiabatically 
    which leads to a completely incoherent pattern
    (Incoherent Merging) due to larger excitations.}
  \label{Incoherence_split_merg}
\end{figure}

Before we proceed to the case of finite-temperature, 
it may be relevant to address the case of a nonadiabatic 
splitting and merging process, i.e., when the double-well potential switches 
on and off too fast. We have carried out simulations 
for switching times as short as 20 ms. In such cases the 
splitting of the condensate (even at $T=0$) becomes completely 
incoherent, there is no trace of phase preservation, and the condensate is
destroyed, as is shown in Fig.~\ref{Incoherence_split_merg}.  
For shorter splitting times, $t_{\rm spl} < 20$ ms,  
it is not even possible to split the Initial BEC into two well 
spatially separated condensates. 
When it comes to the time of merging ($ t_{\rm mer} $), 
a similar behavior is observed if $t_{\rm mer}\le t_{\rm sys}$.

\section {Splitting and merging at finite temperature}
\label{sec4}
So far we have considered only the case of $T=0$ where the initial 
condensate has a well-defined phase. 
In this section, following exactly the same approach as in the 
previous section, we investigate the effects that finite-temperature 
might have on the overall coherence during  the splitting and merging process. 

Fluctuations of phase and density of a BEC are the result of thermal
excitations, and appear usually at finite-temperature~\cite{Petrov3D,Hellweg}.
In such a case the system includes some noncondensed or thermal
particles and hence the total Bose field operator can  be expressed as
\begin{eqnarray}
\hat \Psi (z,t) = \Psi (z,t) + \delta {\hat\Psi(z,t)},
\end{eqnarray}
where $\delta {\hat\Psi}(z,t)$ describes the thermal depletion part.
For a BEC in 3D trapping geometries, fluctuations
of density and phase are only important in a
narrow temperature range near the BEC transition
temperature $T_c$~\cite{Petrov3D}.
For  1D systems, however, phase fluctuations are present at
temperatures far below the degeneracy temperature.
Phase fluctuations can be studied
by solving the Bogoliubov-de Gennes equations 
describing elementary excitations.
Writing the quantum field operator as
$\hat{\Psi}(z)=\sqrt{n_{\rm 1D}(z)}\exp[i\hat{\phi}(z)]$ where  
$n_{\rm 1D}(z)=|\Psi (z,t)|^2$
denotes the condensate density at ${T}=0$ [$n_{\rm 1D}(z=0)=\mu/g_{\rm 1D}$]
the phase and density operators
take, respectively, the following forms~\cite{Shevchenko}:
\begin{equation}
\hat{\phi}(z)=\frac{1}{\sqrt{4 n_{\rm 1D}(z)}}
\sum_{j=1}^{\infty}\Big[f_j^{+}(z)\,\hat{a}_j + 
f_j^{-}(z)\,\hat{a}_j^\dag\Big]
\label{phas-fluc}
\end{equation}
and
\begin{eqnarray}
\hat{n}_{\rm 1D}(z)=
\sqrt{n_{\rm 1D}(z)}\sum_{j=1}^{\infty}i\left(f^{-}_j\,\hat{a}_j - 
f^{+}_j\,\hat{a}^{\dag}_j\right),
\label{dens-fluc}
\end{eqnarray}
where $\hat{a}_j \big( \hat{a}^\dag_j\big)$ 
is the annihilation (creation) operator of the excitations with quantum 
number $j$ and energy $\epsilon_j=\hbar \omega_z \sqrt{j(j+1)/2}$, and  
$f_j^{\pm}=u_j\pm v_j$, where $u_j$ and $v_j$ denote the excitation functions 
determined by the BdG equations. More explicitly, the functions 
$f_{j}^{\pm}$ in a 1D Thomas-Fermi regime and at finite-temperature 
take the form:
\begin{eqnarray}
f_j^{\pm}(z)
&=&\sqrt{\frac{(j+1/2)}{R_{\rm TF}}}
\left(\frac{2\mu}{\epsilon_j}\left[1-
  (z/R_{\rm TF})^2\right]\right)^{\pm 1/2} \nonumber \\
& &\times P_j\left(z/R_{\rm TF}\right)
\label{waveFuns}
\end{eqnarray}
where $P_j(z/R_{\rm TF})$ are Legendre polynomials.
The phase coherence length, in terms of the Thomas-Fermi radius 
$R_{\rm TF}$, is expressed as  $L_{\phi}=R_{\rm TF} T_d\,\mu/{T\hbar\omega_z}$ 
 and characterizes the maximal distance between two phase-correlated points 
in the condensate. Phase fluctuations increase for large trap
aspect ratios and small $N$~\cite{Dettmer}.

Temperature is included at the level of the GPE~(\ref{1D-GPE-T0}) 
by calculating first the density at $T=0$ in the presence 
of the magnetic trap only, 
and then mimicking finite-temperature effects via the phase operator of 
Eq.~(~\ref{phas-fluc}). In other words a phase is imprinted on the condensate 
wave function at this stage. To this aim, we calculate the Bose occupation 
$N_j=\left(e^{\epsilon_j/(k_B\,T)}-1\right)^{-1}$
modes in the Bogoliubov approximation for fixed temperatures replacing the
operators $\hat{a}_j$ and $\hat{a}_j^\dag$ 
by random complex variables $\alpha_j$ and $\alpha_j^\ast$, respectively 
such that $\langle|\alpha_j|^2\rangle=N_j $~\cite{Dettmer}. 

Although the GPE~(\ref{1D-GPE-T0}) in this limit remains valid, the 
BdG equations become modified by the integration over the transverse 
profile of the condensate~\cite{Menotti}. In effect the mode 
functions $f^\pm_j$ are given by Jacobi polynomials, whereas  $\epsilon_j$ 
are given by a slightly different expression than in  the pure 1D case. 
We stress, however, that for the regimes of temperatures we consider, there 
will be no qualitative and practically no quantitative difference between 
the pure- and quasi-1D results.

\begin{figure}[ht!]
  \begin{center}
    \includegraphics[width=\linewidth]{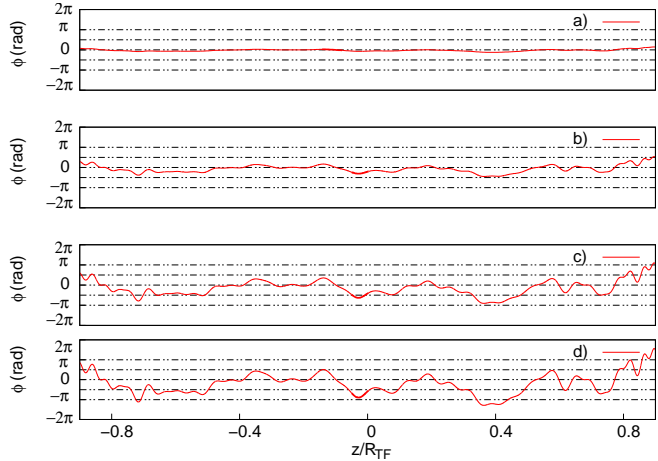}
    \caption{(Color online) Enhancement of phase fluctuations with the 
      increase in temperature. 
      (a) At the top, we have very weak phase fluctuations 
      at very low temperature ($T/T_c=0.01$). 
      (b) With the rise in temperature ($T/T_c=0.1$),  
      the phase fluctuations begin to be enhanced. 
     (c) Still at an intermediate but relatively high temperature 
      ($T/T_c=0.4$), the magnitude of phase fluctuations increases.
     (d) Finally at $T/T_c=0.8$, which is near the critical point, 
       stronger phase fluctuations are displayed.}
    \label{Phas}
  \end{center}
\end{figure}

The dependence of the magnitude of the phase fluctuations
of the condensate on the temperature is shown in Fig.~\ref{Phas}.
Even though we present here results calculated 
for a fixed number of condensate atoms ($N$), 
it is numerically verifiable that the magnitude of the phase fluctuations
 is inversely proportional 
to the square root of the number of atoms and hence to the peak density 
of the condensate as described by Eq.~(\ref{phas-fluc}).
As can be seen by inspecting  the different plots (a)-(d) in Fig.~\ref{Phas}, 
the phase fluctuations get more and more enhanced with the increase of 
temperature. 

\begin{figure}[ht!]
  \begin{center}
    \includegraphics[width=\linewidth]{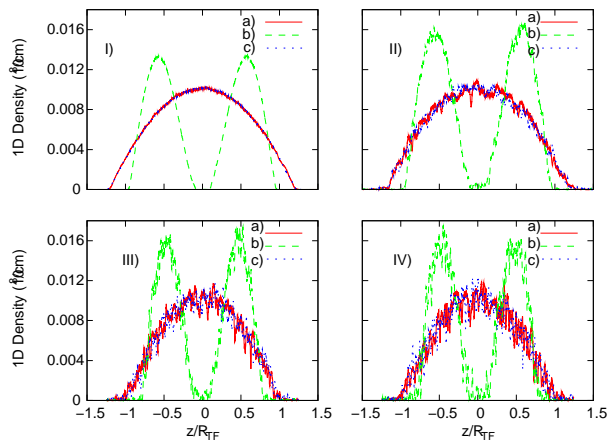}
  \end{center}
  \caption{(Color online) Splitting and  merging of  1D BECs at 
    finite-temperature, i.e., in the presence of phase and 
    density fluctuations. The four plots are the fluctuating densities at 
    $T/T_c$ = (I)  0.01, (II) 0.1, (III) 0.4   and  (IV)  0.8. 
    Each one in turn consists three plots 
    with in it:  curve  {\rm a}, the initial 1D BEC (red solid line);  
    curve {\rm b} the split 1D BECs (green dashed line); and 
    curve {\rm c}, the merged 1D BEC (blue dotted line).}
  \label{DensTemp}
\end{figure}

Having seen the enhancement of phase fluctuations with temperature, we 
proceed now to analyze the coherence properties of splitting and 
merging of a 1D BEC in the presence of these phase fluctuations for 
temperatures in the range of $T_{\phi} < T < T_c$. Our results for  
finite-temperature are summarized in Fig.~\ref{DensTemp}. 

Using exactly the same approach that we employed for the adiabatic 
case at $T=0$ but now including temperature we study again the splitting 
and merging process for the same parameters. 
In this case, the density fluctuations, which are highly suppressed in
elongated 3D condensates, are very pronounced in the 1D density 
after time evolution, when the temperature increases 
from a very small value ($T/T_c=0.01$) 
to a value near the critical point for condensation ($T/T_c=0.8$). 
At high temperatures, the density fluctuations get more and more enhanced. 
However, notice that the effect of fluctuations is very similar on 
the single and on the merged condensates. On the other hand, fluctuations 
on the split condensates remain relatively small due to fact that
the densities in the double wells are higher as shown in all 
plots of Fig.~\ref{DensTemp}. This is in confirmation of  
the prediction of Eq.~(\ref{phas-fluc}). The presence of fluctuations 
on the density profile is a consequence of the BdG equations present in
the GPE equation. In spite of such fluctuations, the split and merged 
condensates present almost the same Thomas-Fermi
density profiles for any temperature $T < T_c$.  

From these observations, we conclude that in spite of the initial 
phase fluctuations in the quasicondensate regime at finite-temperature, 
there is a  preservation of ``phase coherence length'' during the 
splitting and merging process, if adiabaticity is satisfied. 

\section{Summary}
Summarizing, we have discussed coherence effects in the splitting and
merging of a 1D $^{87}$Rb BEC.  This is done by creating two
spatially separated condensates from an initial 1D condensate by
deforming the trapping potential into a double-well potential. We have
analyzed the case of zero temperature as well as finite-temperature in
the so-called quasi-condensate regime, where the phase coherence length
is smaller than the size of the system.  At zero temperature and for a
process adiabatic enough where the splitting and merging times are
much larger than the characteristic time of the system given by 
the inverse of the trap frequency $t_{\rm sys}=2\pi/\omega_z$, 
a coherent splitting followed 
by a coherent merging is achieved and there is a constant relative 
phase between the initial and the final merged condensates. 
On the contrary if the splitting and merging times are not larger than
the relevant time scale, the split condensates acquire a random
relative phases and merging is no longer possible.  
In this case, the
system acquires large density and phase fluctuations on a length scale
comparable with the coherence length $l_c$, and a description based on the
GPE becomes invalid.  In the case of finite-temperature 
our results show that even in the presence of strong phase fluctuations, 
if the process of splitting and merging is carried out fully adiabatically, 
the condensate preserves the Thomas-Fermi density profile and 
there is  phase coherence length preservation.  
In such cases, the merged condensate is a
quasi-condensate with the same initial density profile as the initial
condensate and with the same phase coherence length, although the
``overall'' phase of the quasi-condensate is not preserved.  This
situation occurs even at temperatures very near the critical value for
condensation as long as the trapping potential remains in place.  Our
results may have a useful implication for manipulating 1D BECs at
zero as well as finite-temperatures such as in atom lasers,
interferometry, and solitons. In particular, our results agree
qualitatively well with the recent measurements in Ref.~\cite{Schumm}.

\section*{Acknowledgements}
A.~M. wishes to thank the Deutscher Akademisher Austausch 
Dienst~(DAAD) for financial support. We thank V. Ahufinger, J. Arlt, 
H. Kreutzmann, J. Schmiedmeyer, and 
M. K. Oberthaler for discussions. We acknowledge support from Deutsche 
Forschungsgemeinschaft~(SFB 407, SPP1116, 432 POL, and GK 282), 
from the ESF Programme QUDEDIS, from the EU IP Programme SCALA,
and the Spanish Grant No. MCYT BFM-2002-02588.

\end{document}